# Studies toward a Candidate IDS Neutrino Factory Front-End Configuration

David Neuffer

*Fermilab*
*PO Box 500, Batavia IL 60510*
*January 2010*

**Abstract.** A front end scenario for the IDS neutrino factory is presented. The scenario is based on the Study 2A example for capture, bunching and phase-energy rotation of µ's from a proton source for a neutrino factory, and the goal is the capture of a maximal number of muons in a string of 201.25 MHz rf bunches. We present a candidate release scenario that is somewhat shorter than the Study 2A example and its parameters are optimized for smaller magnetic fields and gradients. We consider the variation of performance with achievable gradient, obtaining acceptable capture with reduced gradients. We also consider variation in production target performance, and develop the specifications toward the practical requirements of the cost study.

## Introduction

For a neutrino factory or a $\mu^+$-$\mu^-$ collider, short, intense bunches of protons are focused onto a target to produce pions, that decay into muons, which are then cooled and accelerated into a high-energy storage ring, where µ decays can provide beams of high-energy neutrinos for a ν-factory[1, 2, 3]. If the $\mu^+$ and $\mu^-$ bunches are counter-rotating and focused to collide in an interaction region, high-luminosity $\mu^+$-$\mu^-$ collisions are possible [4].

In the International Design Study (IDS), the neutrino factory design is being developed with sufficient detail to obtain a cost estimate. The IDS design is based on the International Scoping Study[5] design displayed in Fig. 1. It consists of a high-intensity proton source producing 4MW of ~10GeV protons in short bunches (~2ns rms) at ~50 Hz. These bunches are transported onto a heavy-metal target immersed in a 20T solenoid for production of a large number of π's. The π's are transported into a "Front-End" transport and cooling system that forms the µ's from π-decay into a string of bunches of cooled µ's at ~200 MeV/c. These muons are accelerated through a linac to ~1 GeV, and then through recirculating linacs and an FFAG ring to ~25 GeV, where the µ bunches are transferred into fixed energy rings. µ-decay along the storage ring straight sections provide ν-beams suitable for neutrino oscillation studies.

In this note, we consider the "front-end": the transport from the π-production target through the Buncher, the Phase-Rotator and the Cooling transport, with the goal of developing a candidate scenario for the IDS. In the front end, the pions (and resulting muons) are initially produced within a short bunch length and a broad energy spread. The π's drift from the production target, lengthening into a long bunch with a high-energy "head" and a low-energy "tail", while decaying into μ's. The beam is then transported through a "Buncher" that forms the beam into a string of bunches, and then an "rf Rotator" section that aligns the bunches to (nearly) equal central energies. The μ's are then cooled in a "Cooler" with rf cavities and absorbers. (see figs. 2 and 5) Table I shows baseline parameters for front end solutions, including the ISS scenario, a shorter higher-field version ($N_B = 10$)[6], and a newer lower-field version considered for the IDS. These examples obtain ~0.08 $\mu^+$ and $\mu^-$ per initial 8 GeV proton within the nominal acceptance of a ν-factory.

**Table 1:** Parameters of some buncher/rotator scenarios.

| Simulation cases | | Study 2B ISS baseline | μ-collider example | **Release Candidate (RC)** |
|---|---|---|---|---|
| Parameter | | $N_B=18$ | $N_B=10$ | |
| Bunch spacing number | $N_B$ | 18 | 10 | 10 |
| Drift Length | $L_D$ | 110.7m | 56.4m | 79.6m |
| B-field within drift & buncher | B | 1.75T | 2.0T | 1.5T |
| Buncher Length | $L_B$ | 51m | 31.5 | 33 |
| Buncher rf Gradient | $V_{rf}'$ | 0 to 12 | 0 to 15 | 0 to 9 |
| Buncher rf frequencies | $f_{rf,B}$ | 360→235MHz | 360→240 | 350→232MHz |
| Rotator Length | $L_R$ | 54m | 36m | 42m |
| Rotator Bunch spacing | $N_B+\delta N_B$ | 18.05 | 10.08 | 10.05 |
| Rotator gradient | $V_{rf}'$ | 12.5 | 15 | 12MV/m |
| Rotator rf frequencies | $f_{rf,R}$ | 232 to 202 | 240 to 201.5 | 232 to 202.3MHz |
| Cooler rf Gradient | $V_{rf}'$ | 15 | 15 | 15MV/m |
| μ/ 24 GeV p ($A_T < 0.03$, $A_L < 0.2$m) after rotator. | $\mu/p_{24}$ | 0.126 | 0.124 | 0.11 |
| μ/p ($A_T < 0.03$, $A_L < 0.2$) after LiH cooler | $\mu/p_{24}$ | 0.265 | 0.263 | 0.277 |
| Final transverse emittance | $\varepsilon_{T, rms}$ | 0.0076 | 0.0078 | 0.0067 |
| Final Longitudinal emittance | $\varepsilon_{L, rms}$ | 0.071 | 0.076 | 0.072 |

In these scenarios, the rf cavities in the Buncher and the Rotator are closed-cell pillbox cavities placed within focusing solenoidal fields, with a nominal field of ~2T. (see fig. 3) Recent theoretical models and experimental studies suggest that this configuration enhances the possibility of rf breakdown. The model is that the solenoidal and rf fields would guide and accelerate emitted electrons across the cavity, causing secondary emissions at the opposing surface, and a possible avalanche effect in multiple electron passes. It is considered likely that rf gradient will be limited by the magnetic field with the allowable gradient reduced with increasing magnetic field[7], although it is uncertain what the limits will be for our rf frequencies and B-fields.

There are several potential strategies to circumvent this difficulty which are under consideration:

1. The baseline rf cavities are pillboxes, with Be windows. Open-cell rf cavities may not be as limited, and could be used, but require greater rf power costs.

2. Gas-filled rf cavities are not limited in gradient by magnetic fields, in recent experiments. The energy loss in the gas also provides μ-cooling, although additional gradient is needed to compensate that energy loss. However, there is presently a concern that the electrons produced in the gas may drain rf energy from the cavities[8], and experiments are needed to determine whether that would limit front-end applications.

3. A constant field solenoid can be avoided by using an alternating-solenoid focusing lattice. In the baseline example, the Cooler has an alternating solenoid lattice with a period of 1.5m, obtained by solenoid coils placed between successive rf cavities (see fig. 4). In this configuration the magnetic field lines do not connect directly across the cavity, and rf breakdown may be relatively suppressed. This feature could be enhanced by using a "magnetically-insulated" design, where the cavity surfaces are designed to be parallel to the field lines, preventing accelerated electron paths along the field lines.[9] In a further lattice variation, the rf cavities could be placed in a lattice with very small magnetic fields at the rf locations.[10]

4. The front end can be reconfigured so that the gradient does not exceed $V'_{max}(B)$. The expression for $V'_{max}(B)$ may be dependent on cavity material, temperature and geometry. Palmer has suggested that the lower energy loss and high conductivity of Be may enable much higher gradients than the Cu cavities of the initial designs.[11]

In the present paper, we are maintaining the same general design used for the front end in the ISS, but using field strengths that are somewhat less than those used for the ISS, while being aware of the degree of loss of performance if rf gradients are more limited. Our goal is a design that will still provide relatively good performance with greatly reduced gradients, while providing high performance at its reference parameters. In the following section we present a potential design that may be developed in detail as a release candidate (RC) for the IDS design study based on these considerations.

## Parameters of the RC Design

Our recent studies indicate that high acceptance can be obtained with somewhat reduced focusing fields.[6] Therefore we reduced the baseline focusing field for the Front End to 1.5 T. The overall length of the system was also reduced from the ISS example in order to reduce costs, by about 30%. The reduced length also reduces the μ bunch train length by ~30%. As with the ISS, the front end is matched into a 201.25 MHz ionization cooling system, which is initially set to have ~15MV/m rf cavities and LiH absorbers within an alternating solenoid focusing lattice. The release candidate (RC) is similar to the $N_B=10$ case developed previously but with some differences.

In the RC, the Buncher is $L_B = 33$m in length, and the rf gradient in rf cavities along the buncher increases linearly:

$$V'_{rf}(z) = V'_{B,max}\left(\frac{z}{L_B}\right) \text{ MV/m}.$$

The reference value of $V'_{B,max}$ is 9 MV/m; variations of that value are discussed below. In the baseline configuration, the rf cavities have nominal lengths of 0.5m with pillbox fields, with 0.25m spacing between cavities, and a constant 1.5T solenoidal field is maintained throughout the buncher. The gradual increase of the rf field within the Buncher enables a somewhat adiabatic capture of the muons into strings of bunches at different energies, preparing the bunches for lower-loss acceleration or deceleration in the Rotator.

The rf frequency decreases from 320 to 232 MHz from cavity to cavity along the Buncher. The frequency of the rf cavities in the Buncher is determined by setting two reference muon particles of different momenturm ($P_1$ = 233.5 MeV/c, $P_2$ = 154 MeV/c) at the start of the simulation and requiring that the difference in phase between their passages through the rf cavities be N=10 rf cycles.

The separation of the particles in cτ that develops is given by:

$$\delta(c\tau) = L\left(\frac{1}{\beta_2} - \frac{1}{\beta_1}\right),$$

where δcτ indicates the time delay between the reference particles with speeds $\beta_1$, $\beta_2$, L is the distance from the production target and $\beta_i$ is the reference particle longitudinal speed $v_z/c$. We require that the rf wavelength of the cavity is set to an inverse integer (1/N) of the cτ between reference particles:

$$\lambda_{rf}(L) = \frac{\delta ct_{21}}{N_B} = \frac{L}{N_B}\left(\frac{1}{\beta_2} - \frac{1}{\beta_1}\right)$$

with $N_B$=10 in the present case. With the rf phase = 0 for the reference particles, the rf forms the beam into a string of bunches of different energies.

At the end of the Buncher, the rf program is changed to enable the lower energy bunches to accelerate and higher energy bunches decelerate. The algorithm used for setting this condition is to keep the first reference particle at fixed momentum while the second reference particle is uniformly accelerated through the Rotator section, so as to arrive at the first particle's energy at the end of the Rotator. The phase shift between reference particles is increased slightly ($N_R$ = 10.05) to place the second reference particle at an accelerating phase. This procedure corresponds to using the ACCEL model 12 with phase model 0 in the ICOOL program to set acceleration frequencies and phases. (Previous examples had used phase model 1, in which the first reference particle decelerates while the second accelerates.)

The rf voltage in the rotator cavities was set at $V_R'$ = 12 MV/m, and the rf frequencies of the rf cavities decrease from 232 to 202 MHz over the 42m length of the rotator. The muons continue to form into bunches while the bunches align in momentum (see fig. 5) to ~233.5MeV/c.

At the end of the Rotator the spacing between reference particles is matched to $10\lambda_0$, where $\lambda_0$ is set by the 201.25 MHz rf frequency of the cooling and following acceleration sections. The Cooling section is, initially, a duplicate of the cooling scenario used in the ISS study, consisting of alternating solenoid cells (see fig. 4) of 0.75m length, with 2 LiH absorbers of 1.1cm thickness, and with rf gradients set to 15 MV/m. The capture momentum of ~230 MeV/c is used for cooling and is somewhat higher than the ~210 MeV/c used in the previous examples. While that reduces the rate of transverse cooling, it also reduces longitudinal heating and improves bunching. This allows us to use slightly longer cooling slabs at the same rf gradient. We find optimal cooling with ~1.1cm slabs at 15 MV/m and 1.15cm at 16 MV/m.

The front end capture was simulated using the ICOOL program[12], and results are displayed in table 1. In these simulations an initial population of π's and μ's, as would be produced by 24GeV protons incident on a mercury target at the start of the front end transport, are tracked through the drift buncher, rotator and cooler. In these simulations muons are considered accepted if they meet the ECALC9 criteria of longitudinal amplitudes less than 0.2m and total transverse amplitudes less than 0.03m.[13] The number of muons within the reference acceptance at the beginning and end of the Cooler are displayed in Table 1. The front end was also simulated using an intial beam distribution produced by 8 GeV protons on the target, as generated by C. Yoshikawa using the MARS simulation.[14] These simulations obtain ~1/3 the number of μ's as in the reference 24 GeV p generated sample (~0.08 μ/8 GeV p).

## Variations in RC parameters: Gradients in Buncher and Rotator

In a set of studies, we considered the possibility that the front end will not be able to use rf gradients within magnetic fields that are as large as those in the baseline design. The baseline $N_B=10$ example has rf gradients up to ~12MV/m with solenoidal fields of B=1.5T in the Buncher and Rotator. In the studies we varied the rf voltages in order to study the potential effects of restrictions on maximum rf gradient. We chose a linear ramp in gradient from 0 to 9MV/m for the Buncher, with the Rotator gradient at a constant 12 MV/m, as a reasonable starting point and varied the rf gradients from 15MV/m down to 0 in the Rotator, with similar reductions in the Buncher.

Table 2 summarizes the simulation results of these variations. In these simulations the Buncher and Rotator rf fields are varied and the acceptance at the end of the cooling channel are compared. The cooling channel was fixed at 15 MV/m with 1.1cm LiH. We compare the number of accepted muons at the end of the ~80m cooling channel, using the acceptance criteria discussed above ( $\varepsilon_t$ <0.03m and $\varepsilon_L$<0.2m in ECALC9).

The acceptance is changed very little with Rotator fields varied from 11 to 15 MV/m, and is slightly reduced from ~0.083 to ~0.080, for $V_R$' = 9 to 10 MV/m. It is dramatically reduced for $V_R$' < 7MV/m. It is reduced to ~0.05 for $V_B$'=3, $V_R$'=6, and is reduced to 0.013 if the Buncher and Rotator are turned off. This appears somewhat less sensitive to variations than the previously studied $N_B$ = 10 case,[15] and the RC appears to tolerate lower Buncher / Rotator gradients with good performance.

The rf gradients used here are in a model with 2/3 rf occupancy. If more rf cavities are incorporated, and the drift spaces are filled with rf cavities, the gradients listed could be reduced by 2/3 while obtaining the same performance, and the lower gradient may be more certain to be practical. We thus may consider modifying the scenario by more completely filling the longitudinal space with cavities if necessary.

**Table 2:** µ Capture with varying Bunching and Rotator rf strengths. 8 GeV initial proton beam energies were used in these simulation.

| V' (Buncher) | V' (Rotator) | µ/p at z=245m- ICOOL results |
|---|---|---|
| 0 MV/m | 0 MV/m | 0.013 |
| 3 | 6 | 0.051 |
| 4 | 7 | 0.069 |
| 5 | 8 | 0.075 |
| 6 | 9 | 0.080 |
| 7 | 10 | 0.080 |
| 8 | 11 | 0.0831 |
| 9 (z/L) | 12 | 0.0834 |
| 10 | 13 | 0.0821 |
| 11 | 14 | 0.0839 |
| 12 | 15 | 0.0828 |

# Newer Production Model and Variation

During development of this note, we received a newer initial beam generated by H. Kirk using a more recent MARS simulation model.[16] The model had an 8 GeV proton on a Hg target within 20T or 30T solenoids. The newer model obtains somewhat more µ/p from 8 GeV protons than earlier simulations, and we confirmed that property by inserting that initial beam in place of our previous sample. With the new initial beam the number of muons in the nominal acceptance is increased, by ~10%. (see Table 4) Since this newer model is believed to be more accurate than the earlier case, we have used that beam as the basis for our following studies.

Kirk also considered varying the field of the capture solenoid around the target, with the expectation that a stronger solenoid would enable more capture. He generated an initial beam within a 30T solenoid. We inserted that initial beam into our front end model, with the initial solenoid strength "adiabatically" reduced from 30T to 1.5 T over a length scaled up from our 20 to 1.5 T example. Results are shown in Table 4, and the µ/p within a reference acceptance is increased from ~0.09 to 0.113. The acceptance is reduced to 0.107 if the downstream apertures are reduced to 25cm from the previous nominal 30cm aperture. A higher µ/p can thus be obtained with the higher capture field, by about ~20% as the rf field is increased from 20T to 30T. The 30T magnet will be somewhat more difficult to construct and operate, however.

**Table 3:** μ Capture with Varying Initial Beam and Front End B-field

| Initial Beam | aperture | μ/8GeV p at z=245m |
|---|---|---|
| Baseline from CY[19] | 30cm | 0.083 |
| New 20T beam | 30cm | 0.090 |
| New 30T beam | 30cm | 0.113 |
| New 30T beam | 25cm | 0.107 |

## Other Tests and Studies

The new baseline, as initially developed within the ICOOL model, had various idealizations and not entirely realistic approximations to some of the front end components. We are studying the effects of more realistic models. Initially the solenoidal fields in the buncher and rotator were represented by a constant field 1.5T solenoid. We replaced this by a series of coils designed to obtain ~1.5T fields (using r=65cm, 0.5m long cylindrical coils placed at 0.75m intervals, matching the rf cavity lattice). In the resulting simulations, the capture was reduced by only ~1—2%. The deleterious effect of the finite coil spacing is almost unmeasurable.

In another study we reduced the rf gradient in the Cooler from the nominal 15—16 MV/m to 12 MV/m while reducing the LiH absorbers from 1.1 to 0.8 or 0.85cm. The reduced gradient reduced the acceptance at z=245m from 0.085 to 0.070 in comparable studies. The transverse cooled emittance obtained at that point was $\varepsilon_t$ = 0.0845 rather than 0.071m. If we equalized the cooling, then the 15 MV/m example obtains 0.079 μ/p (after 25m less cooling length). The higher gradient cooling was thus significantly more effective (by almost 20%). From this study it appears that performance is more sensitive to the cooling gradient than to Buncher/Rotator gradient limits.

## rf Parameters and Specifications

In the initial simulations, rf cavities in the Buncher and Rotator were placed at regular intervals (0.5m cavities placed with 0.75m spacing) with each cavity having a different frequency and a gradient matched to the simulation program, and placed within a constant magnetic field. In a more realistic implementation, the cavities will be grouped into a smaller number of frequencies matched to input power sources. For pillbox rf cavities (as used in the MICE experiment [17]) Be windows (or grids) should be added to the ends of each cavity, with cavity parameters rematched to include the energy loss. The constant solenoid field should also be replaced with a more realistic coil configuration. In this section we detail some of the effects of these more realistic rf representations and describe simulation results.

Simulation studies have indicated that reducing the number of independent frequencies by a factor of 3 does not reduce the accepted μ production by more than ~5%.[19] We thus combine the cavities into sets with same frequencies, and tables 4 and 5 show a reduced set of 13 frequencies in the Buncher and 15 in the Rotator that are sufficient for muon collection. The tables show the required rf voltages in each frequency and the resulting rf power requirements.

The baseline rf design is a pillbox cavity and we base the rf parameter specifications on a pillbox model with room temperature Cu walls ($\rho_{Cu}=1.7\times10^{-8}$ ohm-m). The cavity has a radius a and a length L. The cavity skin depth is $\delta=(\rho/\pi f_{rf}\mu_0)^{1/2}$ and the surface resistance is $R_s=\rho/\delta$ ($\delta=4.64\mu$ and $R_s = 3.55$m$\Omega$ at $f_{rf}=200$ MHz). The cavity radius and rf frequency are related by $a=c/(f_{rf}*2.6125)$, which is a=0.574m at 200MHz.[18]

The cavity $Q_0$ is set by the parameters:

$$Q_0 = \frac{2.405 Z_0}{2(\pi f_{rf}\rho\mu_0)^{\frac{1}{2}}(1+\frac{a}{L})}$$

For 200 MHz rf, L=0.5m, $Q_0=5.76\times10^4$. The energy stored in a cavity is:

$$U_0 = \pi\varepsilon_0 L a^2 0.52^2 \frac{E_0^2}{2},$$

where $E_0$ is the rf gradient. At L=.5m, $f_{rf}$ = 200MHz, and $E_0$= 10MV/m, $U_0$= 61.7J. The power dissipated in a cavity is:

$$P_0 = \frac{\pi R_s 0.519^2 E_0^2 a(L+a)}{Z_0^2}$$

At At L=0.5m, $f_{rf}$ = 200MHz, and $E_0$= 10MV/m, $P_0$ = 1.35MW. For the cooling channel we use $E_0$ = 15MV/m, increasing $P_0$ to 3.03MW. Another parameter is the cavity filling time, which is determined by the cavity $Q_0$:

$$T_{fill} = Q_0 \frac{\ln(2.0)}{\pi f_{rf}},$$

This is ~63.5μs at the present parameters.

Another critical parameter is the transit time factor:

$$T_t = \frac{\sin\left(\frac{\pi f_{rf} L}{c}\right)}{\frac{\pi f_{rf} L}{c}}$$

This is $T_t$=0.83 for $f_{rf}$=200 MHz, L=0.5m. For 320MHz this is reduced to 0.59, but restored to 0.73 at $f_{rf}$=320 MHz, L=0.4m. Shorter rf cavities have better transit times, particularly at higher frequencies.

In Table 4, we present the rf frequencies, cavity parameters, and power requirements for the Buncher. For the higher frequency rf cavities (f>300MHz) we reduced the nominal cavity lengths to 0.4m, to avoid unfavorable transit time factors. For the other rf cavities we have chosen cavity lengths of 0.45m. In reducing the number of rf frequencies to 13, we have placed adjacent cavities, at the same frequencies, typically in groups of 3. rf gradients in each cavity are typically ~6MV/m. Table 4 also lists the rf power requirements per cavity. These requirements are grouped into total rf power required per rf frequency (~3 cavities), which assumes that adjacent cavities may share the same power source. The requirements are listed as peak power requirements; approximating the power needed to fill the cavities with some margin (33% more than the peak power dissipation). That peak power is typically a few MW per rf frequency. The duty cycle is ~1% (with ~50Hz pulsing), so the average power is proportionately less.

Table 5 shows rf cavity parameters for the Rotator and the Cooler. The 56 cavities are grouped into 15 separate frequencies from ~230 to ~202 MHz. A reference cavity length of 0.5m and gradient of 12 MV/m is used, and rf power requirements of ~2MW per cavity are obtained. The Cooler rf frequency is set at 201.25 MHz with rf gradients of ~15MV/m in 0.5m rf cavities, and a 75m Cooler would have 100 rf cavities. These cavities would require ~4MW peak power each.

We simulated capture and cooling for a front end with this distribution of cavities and frequencies. In the simulations the Buncher rf cavities had 200μ Be windows at each end and the Rotator rf cavities had 400μ Be windows. The resulting loss of capture was less than ~5%, when compared to a continuous rf model without Be windows. Figure 6 shows μ/p within acceptance criteria in an ICOOL simulation. We obtain ~0.082 μ/p at s=245m within ($\varepsilon_t < 0.03$, $\varepsilon_L < 0.20$) with 15 MV/m cooling. This can be improved somewhat with greater gradient and cooling; with 1.2cm LiH slabs and 17MV/m rf we obtain ~0.089 μ/p within the same aperture criteria. See figs. 7 and 8.

This demonstrates that the rf can be grouped into a manageable number of rf frequencies, and the collection and cooling can be developed within a practical set of parameters. The rf and magnetic fields can be specified to a level that a first-order cost and practicality study can be implemented. While it is not yet certain what precise gradient/fields should be used, the RC parameters presented here can be useful as a basis in setting the scale for a neutrino factory front end of this type. The IDS study can then determine whether that scale is indeed practical, and designs based on the IDS and supporting rf and magnetic field experiments can be used for construction of the future neutrino factory and muon collider factory.

## Acknowledgments

I thank C. Yoshikawa, J. Gallardo, R. Palmer, C. Rogers and R. Fernow for important contributions and discussions. Research supported by Department of Energy under contract no. DEAC02-07CH11359.

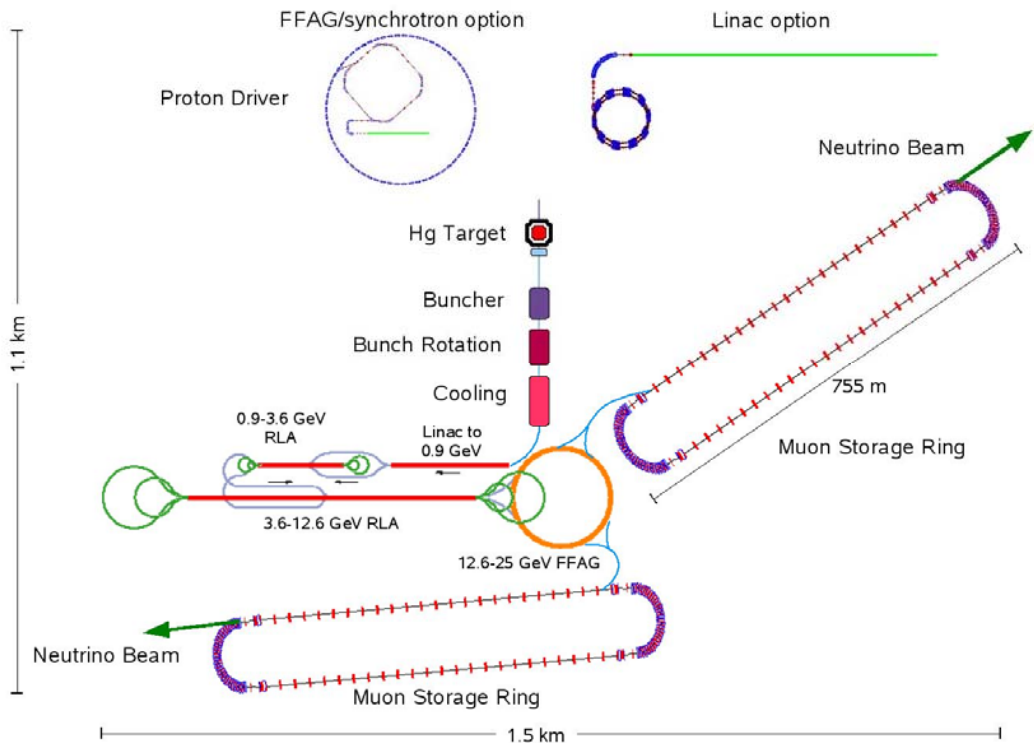

**Figure 1:** Overview of the International Scoping Study Neutrino Factory. It consists of a Proton Driver that produces intense proton bunches (FFAG and Linac based versions are shown), a Mercury target with Buncher, Bunch Rotation and Cooling to obtain intense muon bunches, a 25 GeV muon accelerator consisting of a Linac + 2RLA's (Recirculating Linear Accelerators) + an FFAG, and storage rings, in which muon decays provide neutrino beams for long-baseline neutrino detectors.

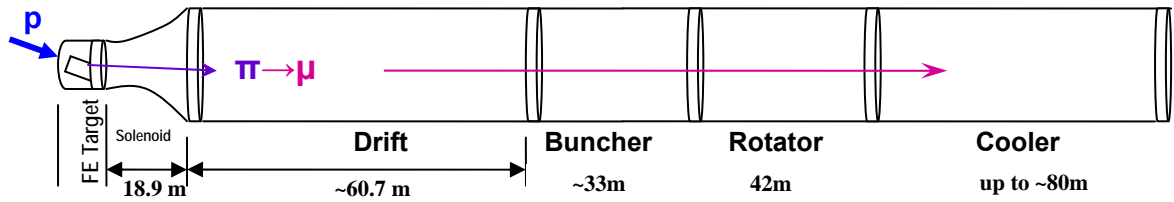

**Figure 2.** Schematic view of the components of the candidate IDS front-end system, showing an initial drift (18.9 +60.7m) from the production target, the Buncher (33m), and the phase-energy (ϕ-δE) Rotator (42m) leading into a Cooling section of up to ~80m. π's would be produced by protons on a target at the beginning of the drift, and decay to μ's in the drift, while lengthening in phase. The Buncher and ϕ-δE Rotator form the μ's into a string of bunches matched into the Cooler.

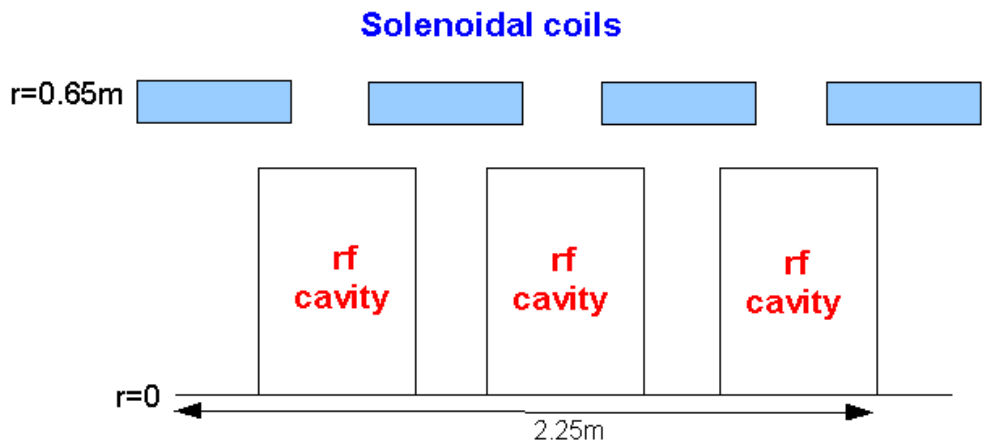

**Figure 3:** Baseline configuration for the buncher and phase-energy rotation sections of the ν-factory front end. The sections consist of rf cavities (cylindrical "pill-boxes") that are spaced at 0.75m intervals with a nominal length of 0.5m (0.25m between cavities). The cavities are placed within a constant-field solenoid, with B=1.5T.

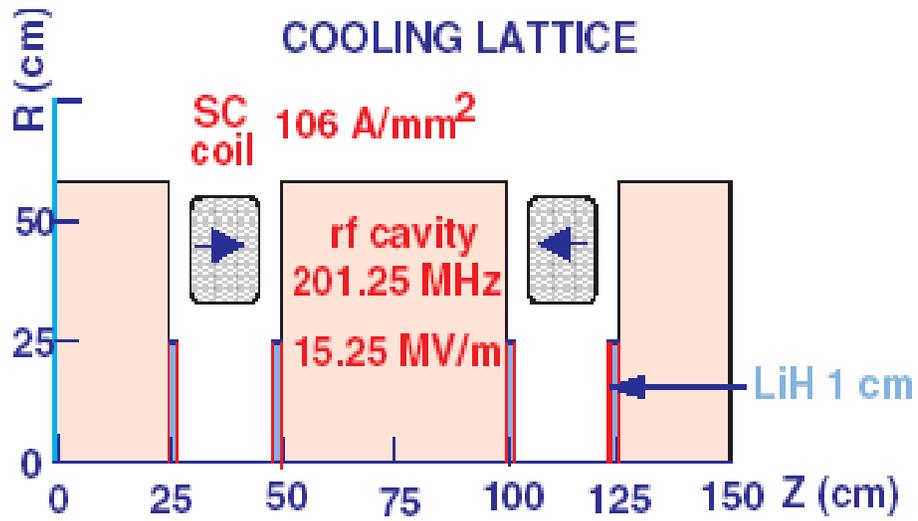

**Figure 4:** Cooling cell from Study 2B.[3] The cooling cell includes 2 rf cavities, 4 LiH absorbers and two superconducting coils. An alternating solenoid field is obtained from the coils, with maximum on-axis fields of ~±2.8T.

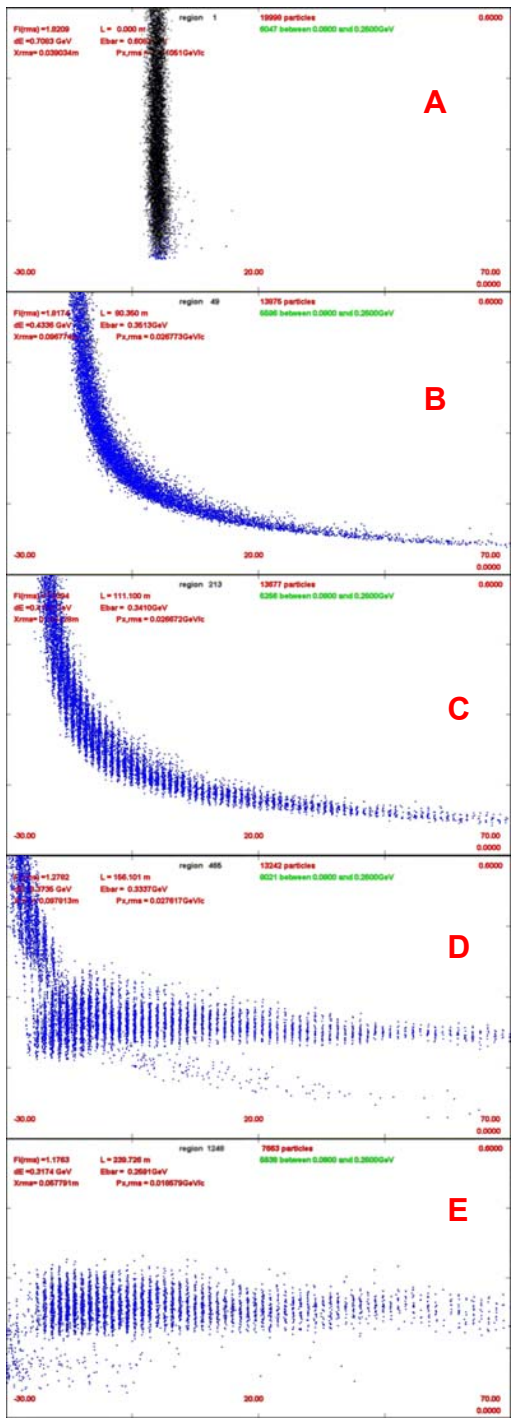

**Figure 5:** ICOOL simulation results of the buncher and phase rotation, at the parameters of the $N_B$=10 RC scenario described in the text. A: π's and μ's as produced at the end of a 1.0m long target. B: μ's at z=80m after drift. C: μ's at z=111m, the end of the buncher. The beam has been formed into a string of ~200MHz bunches at different energies. D: At z= 156m after φ-δE rotation; the bunches are aligned into nearly equal energies. E: At z=236m, after ~75m of ionization cooling cells. In each plot the vertical axis is momentum (0 to 0.60 GeV/c) and the horizontal axis is longitudinal position with respect to a reference particle (-30 to 70m).

**Table 4 Buncher rf parameters**

| RF frequency | Total voltage | cavities | Gradient | $P_0$ rf power/cavity-thermal | Peak rf Power /rf frequency |
|---|---|---|---|---|---|
| 319.63 | 1.368 | 1 (0.4m) | 4 MV/m | ~0.12 MW | 0.2 MW |
| 305.56 | 3.915 | 2 (0.4m) | 5MV/m | 0.20 | 0.6 |
| 293.93 | 3.336 | 2 (0.45m) | 4 MV/m | 0.14 | 0.5 |
| 285.46 | 4.803 | 2 (0.45m) | 5.5MV/m | 0.27 | 1.0 |
| 278.59 | 5.724 | 2 (0.45m) | 6.4 MV/m | 0.38 | 1.25 |
| 272.05 | 6.664 | 3 (0.45m) | 5MV/m | 0.24 | 1.5 |
| 265.80 | 7.565 | 3 (0.45m) | 5.7MV/m | 0.31 | 1.5 |
| 259.83 | 8.484 | 3 (0.45m) | 6.5MV/m | 0.42 | 2.0 |
| 254.13 | 9.405 | 3 (0.45m) | 7MV/m | 0.49 | 2.25 |
| 248.67 | 10.326 | 4 (0.45m) | 6MV/m | 0.37 | 2.25 |
| 243.44 | 11.225 | 4(0.45m) | 6.5MV/m | 0.44 | 2.5 |
| 238.42 | 12.16 | 4 (0.45m) | 7MV/m | 0.52 | 3.0 |
| 233.61 | 13.11 | 4 (0.45m) | 7.5MV/m | 0.58 | 3.5 |
|  | 98.085 | 37 cavities |  |  |  |

**Table 5: Rotator and Cooler rf parameters**

| RF frequency | Total Voltage | Cavities | rf Gradient | $P_0$ MW/cavity | Peak Power/cavity |
|---|---|---|---|---|---|
| 230.19 | 18 | 3 (0.5m) | 12MV/m | 1.68 | 2.25 MW |
| 226.13 | 18 | 3 (0.5m) | 12MV/m | 1.71 | 2.3 |
| 222.59 | 18 | 3 (0.5m) | 12MV/m | 1.74 | 2.35 |
| 219.48 | 18 | 3 (0.5m) | 12MV/m | 1.76 | 2.35 |
| 216.76 | 18 | 3 (0.5m) | 12MV/m | 1.78 | 2.4 |
| 214.37 | 18 | 3 (0.5m) | 12MV/m | 1.80 | 2.4 |
| 212.28 | 18 | 3 (0.5m) | 12MV/m | 1.82 | 2.45 |
| 210.46 | 18 | 3 (0.5m) | 12MV/m | 1.84 | 2.45 |
| 208.64 | 24 | 4 (0.5m) | 12MV/m | 1.85 | 2.5 |
| 206.90 | 24 | 4 (0.5m) | 12MV/m | 1.86 | 2.5 |
| 205.49 | 24 | 4 (0.5m) | 12MV/m | 1.88 | 2.5 |
| 204.25 | 30 | 5(0.5m) | 12MV/m | 1.90 | 2.55 |
| 203.26 | 30 | 5(0.5m) | 12MV/m | 1.91 | 2.55 |
| 202.63 | 30 | 5(0.5m) | 12MV/m | 1.92 | 2.55 |
| 202.33 | 30 | 5(0.5m) | 12MV/m | 1.92 | 2.55 |
| Rotator total | 336 | 56 |  |  |  |
| 201.25 (Cooler) | 750 | 100(0.5m) | 15MV/m | 3.01 | 4 MW |

**Figure 6.** Simulation results for muon collection in the RC front end, with realistic rf cavities and frequencies (13 in Buncher, 15 in Rotator), and Be windows. The three lines show muons within acceptance criteria in the channel of up to 260m length. The upper (dark blue) trace shows all (positive) muons within a 100 to 350 MeV/c acceptance, obtained from a reference set of 10000 8GeV protons at the target (z=0). The light blue trace shows muons accepted within 201.25 MHz rf buckets with amplitudes $\varepsilon_t < 0.03$ and $\varepsilon_L < 0.2$. This accepted subsample has an rms emittance of ~0.004 m (normalized). The number of accepted µ's increases from ~0.038µ/p at s=155m (beginning of 201.25MHz cooling) to ~0.081µ/p at s=240m. The lower curve (purple) shows muons with amplitudes $\varepsilon_t < 0.015$ and $\varepsilon_L < 0.2$.

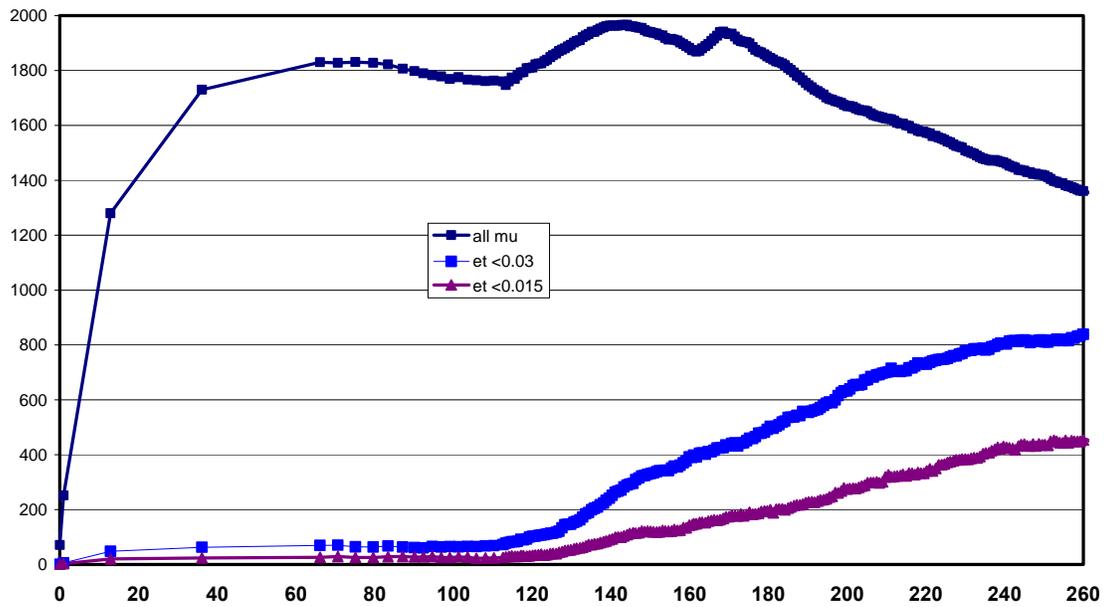

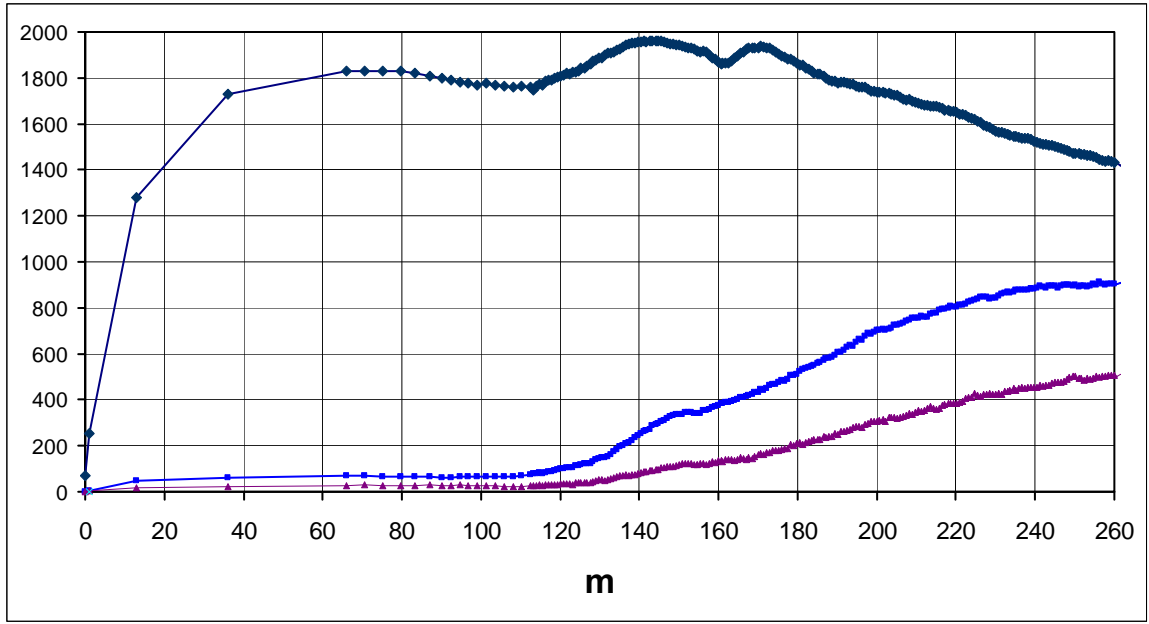

**Figure 7.** Simulation results for muon collection in the RC front end, with realistic rf cavities and frequencies (13 in Buncher, 15 in Rotator), and Be windows. This is the same case as figure 6, except that the cooling rf has been changed to 17MV/m with the absorber lengths increased to 1.2cm. The number of accepted μ's increases from ~0.038μ/p at s=155m (beginning of 201.25MHz cooling) to ~0.09μ/p at s=240m. The lower curve (purple) shows muons with amplitudes $\varepsilon_t$ < 0.015 and $\varepsilon_L$ <0.2.

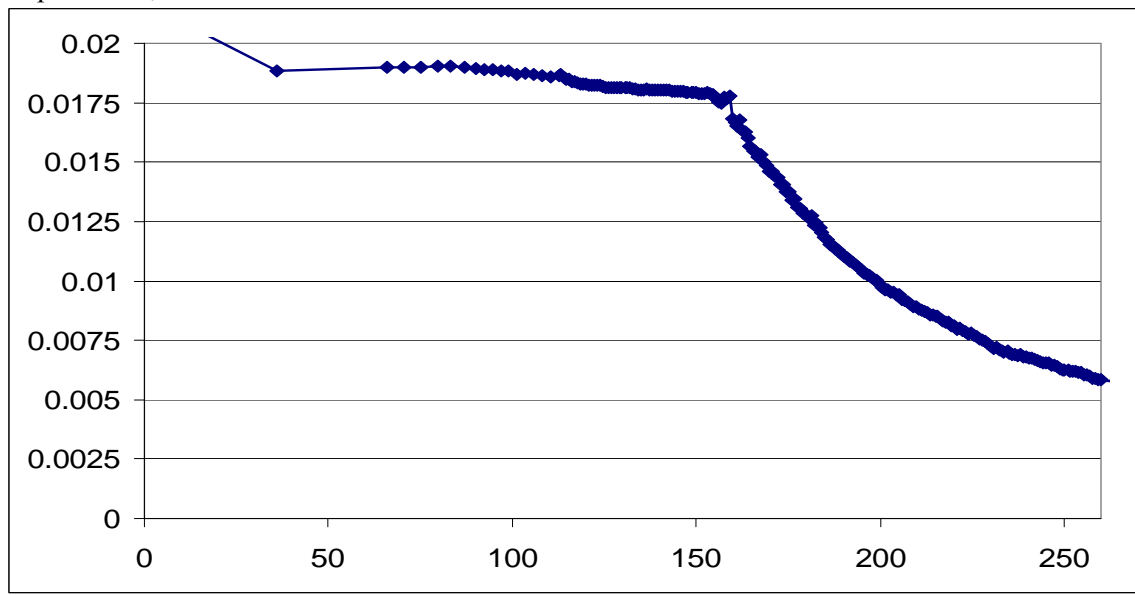

**Figure 8.** Transverse rms emittance through the cooling system used for figure 7, as calculated using ECALC9 from an ICOOL simulation. Transverse cooling within the Buncher and Rotator (from Be windows) reduces transverse emittance from ~0.019m (normalized, rms) to ~0.0175m. Ionization cooling in the cooler reduces the rms emittance to ~0.006m after 100m of cooling (2.4cm LiH per 0.75m cell.).